\pdfoutput=1

\documentclass[11pt]{article}

\usepackage[final]{acl}
\usepackage{times}
\usepackage{latexsym}
\usepackage[T1]{fontenc}
\usepackage[utf8]{inputenc}
\usepackage{inconsolata}
\usepackage{graphicx}
\usepackage{subfigure}
\usepackage{paralist}
\usepackage[normalem]{ulem}
\usepackage{multirow}
\usepackage{amsmath}
\usepackage{listings}
\usepackage{makecell}

\usepackage[inkscapelatex=false]{svg}

\usepackage{amsmath,amssymb,amsfonts}
\usepackage{algorithmic}
\usepackage[linesnumbered,ruled]{algorithm2e}

\usepackage{textcomp}
\usepackage{array}
\usepackage{framed}
\usepackage{caption}
\usepackage{xcolor}
\usepackage{url}
\usepackage{enumitem}
\usepackage{algorithm2e}

\usepackage{booktabs}
\usepackage{color}
\usepackage{tikz}
\usepackage{setspace}

\newcommand{\system}{\textit{\sloppy{SelRepair}\@}}
\newcommand{\tabincell}[2]{\begin{tabular}{@{}#1@{}}#2\end{tabular}}

\newcolumntype{C}[1]{>{\centering\arraybackslash}p{#1}}

\title{Accelerating Automatic Program Repair with Dual Retrieval-Augmented Fine-Tuning and Patch Generation on Large Language Models}

\author{
 \textbf{Hanyang Guo\textsuperscript{1,2}},
 \textbf{Xiaoheng Xie\textsuperscript{3}},
 \textbf{Hong-Ning Dai\textsuperscript{2}\thanks{Corresponding author.}},
 \textbf{Peng Di\textsuperscript{3}},
 \textbf{Yu Zhang\textsuperscript{3}},
 \textbf{Bishenghui Tao\textsuperscript{4}},
 \textbf{Zibin Zheng\textsuperscript{1}}
\\
 \textsuperscript{1}School of Software Engineering, Sun Yat-sen University,\\
 \textsuperscript{2}Department of Computer Science, Hong Kong Baptist University,\\
 \textsuperscript{3}Ant Group \\
 \textsuperscript{4}School of Science and Technology, Hong Kong Metropolitan University
\\
 {
   \texttt{guohy36@mail2.sysu.edu.cn,xiexie@antgroup.com,hndai@ieee.org}} \\{\texttt{dipeng.dp@antgroup.com,zhzibin@mail.sysu.edu.cn}
}
}

\begin{document}
\maketitle
\begin{abstract}

Automated Program Repair (APR) is essential for ensuring software reliability and quality while enhancing efficiency and reducing developers' workload. Although rule-based and learning-based APR methods have demonstrated their effectiveness, their performance was constrained by the defect type of repair, the quality of training data, and the size of model parameters. Recently, Large Language Models (LLMs) combined with Retrieval-Augmented-Generation (RAG) have been increasingly adopted in APR tasks. However, current code LLMs and RAG designs neither fully address code repair tasks nor consider code-specific features. To overcome these limitations, we propose \system{}, a novel APR approach with integration of a fine-tuned LLM with a newly-designed dual RAG module. This approach uses a bug-fix pair dataset for fine-tuning and incorporates semantic and syntactic/structural similarity information through an RAG selection gate. This design ensures relevant information is retrieved efficiently, thereby reducing token length and inference time. Evaluations on Java datasets show \system{} outperforms other APR methods, achieving 26.29\% and 17.64\% in terms of exact match (EM) on different datasets while reducing inference time by at least 6.42\% with controlled input lengths.
\end{abstract}

\section{Introduction} \label{introduction}

Program Repair (PR) is the process of identifying and correcting errors (often called bugs) in a software program by using automated tools or manual techniques with an aim to improve the software's reliability and performance~\cite{Urli2018}. 
PR is a time-consuming and labor-intensive task,  e.g., fixing bugs taking up more than 1/3 of the software maintenance time~\cite{Lientz1978} and 90\% of software maintenance cost~\cite{britton2012quantify}. To improve the efficiency of PR, automatic program repair (APR)~\cite{Goues2021} has been proposed to reduce time consumption and human efforts. 
Some APR approaches, such as heuristic-based approaches~\cite{Westley2009}, template-based approaches~\cite{Meng2023}, and semantics-driven approaches~\cite{Nguyen2013} are limited by the manual process (involving laborious heuristic rule design and template design) and the types of bugs fixed. Although deep-learning-based APR can address the limitations related to the bug types, its performance is mainly influenced by the model parameters and the quality of the training data~\cite{Wang2023}.

Considering the limitations of conventional APR approaches, LLMs have recently been proposed to complete APR tasks~\cite{Huang2023} 
owing to their stronger natural language understanding and even code understanding capability obtained by extensive training on vast amounts of corpus. Nowadays, there are two ways of adopting LLMs to complete APR tasks~\cite{soylu2024}: \textit{prompt engineering} and \textit{fine-tuning} (refer to Appendix~\ref{sec:taxonomy} for more details). As for prompt engineering, since most popular generalized LLMs do not include APR-related pre-training tasks, it is difficult to design an ideal set of prompts to target generic APR tasks. Regarding fine-tuning approaches, most of them are adopted on LLMs with fewer than 1B parameters. For models with more than 1B parameters, the primary fine-tuning method is \underline{P}arameter-\underline{E}fficient \underline{F}ine-\underline{T}uning (PEFT) fine-tuning, which \emph{cannot fully unleash the potential of LLMs in APR}. In addition, for both prompt engineering and fine-tuning, the design of most prompts includes natural language contexts, such as issue/error descriptions and function requirements. Although this kind of prompt can provide additional details to understand codes, it also \emph{increases the prompt complexity and limits the usage scenarios}. Specifically, natural language descriptions are redundant to some simple syntax errors or common PR tasks, thereby easily causing the prompt to exceed the length limit~\cite{Chen2024}. Moreover, prompts with natural language descriptions cannot handle those scenarios, in which developers or students provide error codes without detailed error descriptions at the initial stages of program development. 

Recently, \underline{R}etrieval-\underline{A}ugmented \underline{G}eneration (RAG) has been adopted to improve the performance of LLMs. RAG generates accurate outputs by firstly retrieving relevant information (usually from an external specialized knowledge base) and then feeding this retrieved information into LLMs as contexts, thereby greatly enhancing the ability of LLMs (Appendix~\ref{sec:RAG example} depicts an example of how RAG contributes to APR). In the LLM-based APR task, RAG has been utilized to both prompt engineering~\cite{Nashid2023} and fine-tuning~\cite{Wang2023} modules. However, existing RAG approaches only adopt code semantics similarity while overlooking other code features, such as code structure and syntax information. The utilization of key code features needs a fine-grained program analysis while few approaches leverage RAG based on diverse code features. More importantly, these RAG-based approaches lack the \textit{validation of RAG selection} since they do not judge the necessity of RAGs for APR tasks. The lack of judging RAGs may result in redundant information being added to the input, thereby \emph{increasing the model inference time} and \ textit{degrading performance}. 

In order to fill the above gaps, we propose \system{}, a novel \underline{Sel}ective RAG-based program \underline{Repair} framework by full-parameter fine-tuning LLM. This framework considers \textit{both semantics and syntax information} matching for retrieval based on buggy codes. Moreover, a newly-designed \textit{RAG selection gate} is adopted to determine the necessity of RAGs by setting a threshold to determine whether the extracted bug-fix pair needs to be added to the context. The RAG selection gate can achieve efficient retrieval and controlled prompt length, thereby decreasing inference time. 
Further, we utilize existing APR datasets and design a \textit{code-only} prompt to \textit{full-parameter fine-tune} a large-parameter code LLM 
for APR tasks. As a result, the capabilities of LLMs have been fully exploited for APR tasks. The code-only prompt can be applied to diverse scenarios while controlling the prompt length. We evaluate our method by conducting extensive experiments on a public APR dataset of Java and an enterprise dataset. The experimental results demonstrate that integrating full-parameter fine-tuned LLMs with dual RAG greatly contributes to outstanding APR performance in terms of Exact Match (EM) and CodeBLEU.

The contributions of this paper are fourfold:
\begin{compactitem}
\item We propose \system{}, an APR framework that leverages similar bug-fix pairs as the context and fine-tuned LLM to achieve better PR performance than other SOTA approaches.
\item In order to extract the unique features of codes, we construct a \textit{dual RAG module} considering not only semantics similarity but also syntax as well as structure similarity to retrieve relevant context for APR. Both the retrievals contribute to the superior performance of \system{}. 
\item To ensure the effective validation of RAG, we design a \textit{RAG selection gate} to determine whether the extracted information is input into the LLM as a context. With the utilization of the RAG selection gate, we control the average input length, which decreases to 60.53, 133.16, and 992.25 in Java datasets (with two different code lengths) and a C/C++ dataset, respectively, while the inference time decreases by 6.42\%, 13.77\%, and 9.95\%, respectively. 
\item We utilize a \textit{code-only} prompt for APR tasks and adopt it to full-parameter fine-tune LLM, thereby fully exploiting the capabilities of LLMs and making the prompt concise to apply to diverse scenarios. Our approach outperforms other state-of-the-art approaches in a public APR dataset and an enterprise dataset. It can achieve 26.29\%, 17.64\%, and 25.46\% of EM in Java datasets  (with two different code lengths) and a C/C++ dataset, respectively. It can also generate 59 correct patches in the enterprise dataset.
\end{compactitem}

\section{Related Work}

\subsection{Automatic Program Repair}

As mentioned in \S~\ref{introduction}, conventional APR approaches can be categorized into the following four types. 
\textbf{Heuristic-based approaches} adopt heuristic rules or genetic algorithms to generate patches such as \textit{GenProg}~\cite{Goues2012}, Marriagent~\cite{Kou2016}, pyEDB~\cite{Fatmah2014}. 
\textbf{Template-based approaches} use predefined fix templates to guide code modifications~\cite{Meng2023}. Typical template-based approaches include \textit{TBar}~\cite{Liu2019} and \textit{PAR}~\cite{Kim2013}. 
\textbf{Semantics-driven approaches} such as \textit{SemFix}~\cite{Nguyen2013} use symbolic execution and test suites to extract semantic constraints, and then synthesize repairs satisfying the extracted constraints by program synthesis~\cite{Le2018}.
Considering the fix-type limitations of the above APR methods, \textbf{deep-learning-based approaches} have kept rapidly evolving by adopting neural machine translation techniques in natural language processing to generate repair patches~\cite{Tufano2019,Jiang2021,Rahul2017}. 

\begin{figure*}[t]
  \centering
  \includegraphics[width=\linewidth]{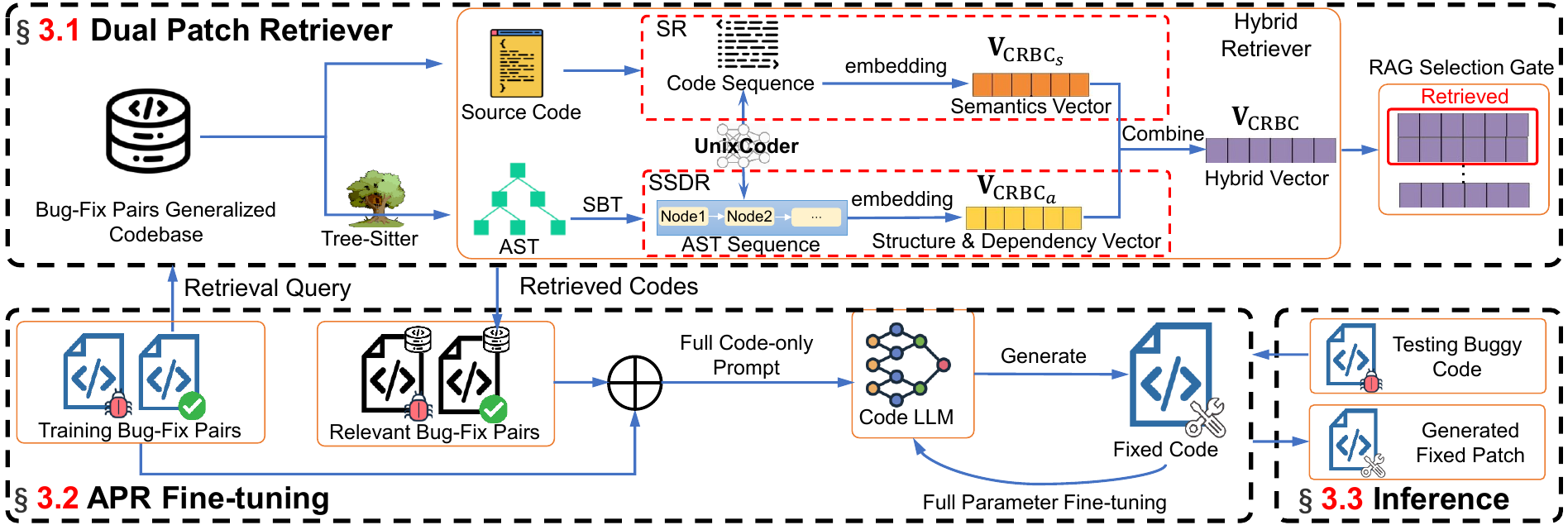}
  \caption{The Workflow of \system{}}
  \label{Overview}
\end{figure*}

Since deep-learning-based approaches are limited by model parameters and training-data quality, \textbf{LLM-based approaches} have been proposed for APR~\cite{Zhang2023, Xia2022}. Specifically, prompt-engineering-based methods extract knowledge by static analysis tools and combine the knowledge with buggy codes to construct prompts~\cite{Pearce2023}. Fine-tuning-based methods, such as \textit{RAP-Gen}~\cite{Wang2023} adopt code-only prompts to fine-tune LLM. Some other approaches use PEFT fine-tuning and RAG for APR~\cite{silva2024repairllama} or APR assistance~\cite{li2024}.
In our research, we adopt full-parameter fine-tuning on an LLM with larger parameter sizes and optimize RAG by using the selection gate and dual retrievals.

\subsection{LLM for SE tasks}

With the rapid development of LLMs, many LLMs have been proposed to be adopted in software engineering (SE) tasks~\cite{zheng2023,zheng2024}. On the one hand, some studies utilized prompt engineering on generalized LLMs~\cite{minaee2024}. such as code summarization~\cite{sun2023} and vulnerability detection~\cite{zhou2024}. On the other hand, some specialized LLMs (i.e., code LLM) such as \textit{CodeBERT}~\cite{feng2020}, \textit{GraphCodeBERT}~\cite{guo2021graphcodebert}, and \textit{CodeT5}~\cite{wang2021} were proposed in the SE field. Since these LLMs mainly use programming languages as the pre-training corpus and SE tasks as the pre-training tasks (e.g., code completion and identifier prediction), these models perform better than generalized LLMs in some complex SE tasks~\cite{Chen2023}. Therefore, some studies adopted these models for specific SE tasks, such as vulnerability detection~\cite{WANG2024} or code search~\cite{Wang2023-1}. 

Recently, some models with larger parameter sizes (more than 1 billion parameters), e.g., \textit{CodeLLama}~\cite{Baptiste2024} and \textit{StarCoder 2}~\cite{lozhkov2024} have been proposed, although the research related to full-parameter fine-tuning LLMs is still relatively limited. These LLMs typically include \textit{BASE} version and \textit{INSTURCT} version. The \textit{INSTURCT} version is the fine-tuned model by using specific natural language instructions.  Instead of using \textit{INSTURCT}, we only adopt the \textit{BASE} version that utilizes code information as a pre-training corpus. The goal of this paper is to utilize RAG and full-parameter fine-tuning on code LLMs for APR.

\section{Methodology} \label{sec:method}
Figure~\ref{Overview} depicts the workflow of the proposed \system{}. We design a dual patch retriever considering both semantics and structure-dependency information. An RAG selection gate is also added in the dual patch retriever (in \S~\ref{RAG}). Then, we adopt the retriever to get relevant bug-fix pairs as context and combine them with buggy code as code-only prompts to fine-tune code LLMs 
for APR (in \S~\ref{fine-tuning}). At last, we utilize the fine-tuned models to generate fixed code (in \S~\ref{inference}).

\subsection{Dual Patch Retriever} \label{RAG}
\textbf{Codebase Construction.}
RAG can enhance the ability of LLM since it can provide expert knowledge that contributes to the task. We construct a codebase that includes APR-related knowledge. Specifically, the codebase consists of existing method-level buggy codes and their corresponding fix patches (i.e., bug-fix pairs). Our goal is to retrieve the relevant bug-fix pairs as the context. In contrast to most existing methods, where the codebase built on top of \emph{repository-level} only retrieves \emph{repository-level contexts}~\cite{Zhang2024, xia2024}, we construct a \emph{generalized codebase} based on \textit{across repositories}.

\textbf{Hybrid Retriever.} 
The dual patch retriever is an RAG module that aims to retrieve the most relevant bug-fix pairs as the context. Among those studies related to LLM for SE tasks, some RAG modules have been utilized to retrieve relevant information. They adopt similarity metrics, such as BM25 and code embedding~\cite{Wang2023, Nashid2023} to get the most relevant code. BM25 considers the code token frequency as the relevance metric while code embedding converts the code to vectors for similarity calculation. However, these two features only consider the source code information as a reference for relevant information, though the source code only contains superficial semantics without including other programming language features, such as syntax information, variable type information, control flow information, and so on. To tackle this issue, we introduce \underline{a}bstract \underline{s}yntax \underline{t}ree (AST), an abstract representation of the syntactic structure of source code to attain complete information for retrieval. AST is represented by a tree structure, in which each node has both \textit{type} and \textit{value} information. \textit{Type} indicates the role of the node in the syntactic structure, such as \texttt{\small if\_statement} and \texttt{\small formal\_parameters}, etc., while \textit{value} indicates specific code information, which is consistent with the source code information. By adopting ASTs, we can attain static structures.
Based on the above information, we can also get the data dependency. Both source code and AST can construct complete program information, thus increasing the retrieval reliability. Considering both semantics and static structures of programs, we construct a hybrid retriever by combining a \textbf{\underline{s}emantics \underline{r}etriever (SR)} and a \textbf{\underline{s}tatic \underline{s}tructure and \underline{d}ependency \underline{r}etriever (SSDR)}.

Algorithm~\ref{algHR} (in Appendix~\ref{app:hybrid}) elaborates on the working procedure of the hybrid retriever. We aim to retrieve the most relevant bug-fix pairs from the codebase. Firstly, in line~\ref{line:ASTParse}, we adopt \texttt{\small AST\_Parse} function to get the AST of the buggy code to fix (i.e., target buggy code). In \texttt{\small AST\_Parse}, we conduct an incremental grammar parsing library for parsing programming languages called Tree-Sitter~\cite{Latif2023} to generate the AST. In order to extract features of both code and AST, we utilize a code pre-trained model, \textit{UnixCoder}~\cite{guo2022} to convert code and AST into semantics vector and structure vector, respectively. We use \textit{UnixCoder} mainly because it adopts both code and AST as the training corpus in pre-training tasks including Masked Language Modeling, Unidirectional Language Modeling, and Code Fragment Representation Learning. Thus, we do not need to fine-tune or retrain an embedding model of AST. 
Since \textit{UnixCoder} takes a sequence as input, we traverse the source code to obtain a code sequence as \textit{UnixCoder}'s input. It is necessary to traverse AST to a flattened sequence for model understanding (line~\ref{line:ASTtraversal}) due to AST's tree structure. Referring to~\cite{guo2022}, we transform AST nodes to the sequence (refer to Appendix~\ref{sec:AST traversal} for more details).  After obtaining the sequences of both codes and ASTs, we input them to \textit{UnixCoder} and get the source code vector $\mathbf{V}_\mathrm{BCs}$ and the AST vector $\mathbf{V}_\mathrm{BCa}$ of target \underline{b}uggy \underline{c}ode ($\mathrm{BC}$) (lines~\ref{line:BCc}-\ref{line:BCa}). Then, we calculate the average vector $\mathbf{V}_\mathrm{BC}$ to get the target buggy code's hybrid feature vector.

In the retrieval phase, we iterate each bug-fix pair in the codebase (line~\ref{line:iterate}) and get the \underline{c}andidate \underline{r}elevant \underline{b}uggy \underline{c}ode ($\mathrm{CRBC}$) and \underline{c}andidate \underline{r}elevant \underline{f}ixed \underline{c}ode ($\mathrm{CRFC}$). Similarly, we adopt Tree-Sitter to parse the AST and \texttt{\small AST\_traversal} to get the AST sequence of $\mathrm{CRBC}$ (lines~\ref{line:ASTParseCR}-\ref{line:ASTtraversalCR}). \textit{UnixCoder} is also adopted to get the source code vector $\mathbf{V}_\mathrm{CRBCs}$ and the AST vector $\mathbf{V}_\mathrm{CRBCa}$ (lines~\ref{line:CRBCc}-\ref{line:CRBCa}). The hybrid feature vector of $\mathrm{CRBC}$ $\mathbf{V}_\mathrm{CRBC}$ is also attained (lines~\ref{line:CRBC}).

In the hybrid retriever, we use the cosine similarity of the hybrid feature vectors to measure the relevance between the target buggy code and the bug-fix pair in the codebase (line~\ref{line:CosSimilarity}). The cosine similarity can be calculated as follows:
\begin{equation} \small
   \kappa(\mathbf{V}_\mathrm{CRBC}, \mathbf{V}_\mathrm{BC}) = \frac{\mathbf{V_\text{CRBC}} \cdot \mathbf{V_\text{BC}}}{\|\mathbf{V_\text{CRBC}}\| \times \|\mathbf{V_\text{BC}}\|},
\end{equation}
where $\mathbf{V}_\mathrm{CRBC}$ represents the hybrid feature vector of $\mathrm{CRBC}$ in the bug-fix pair and $\mathbf{V}_\mathrm{BC}$ represents the hybrid feature vector of the $\mathrm{BC}$ to be fixed. The term $\mathbf{V}_\mathrm{CRBC}$ $\cdot$ $\mathbf{V}_\mathrm{BC}$ represents the dot product, which is calculated by
$\mathbf{V}_\mathrm{CRBC} \cdot \mathbf{V}_\mathrm{BC} = \sum_{i=1}^{n} V_{\mathrm{CRBC}_i} V_{\mathrm{BC}_i}$,
where $V_{\mathrm{CRBC}_i}$ and $V_{\mathrm{BC}_i}$ are the $i^\text{th}$ elements of $\mathbf{V}_\mathrm{CRBC}$ and $\mathbf{V}_\mathrm{BC}$ vectors, respectively, with $n$ vector dimension. Vectors $\mathbf{V}_\mathrm{CRBC}$ and $\mathbf{V}_\mathrm{BC}$ have norms denoted by $\|\mathbf{V}_\mathrm{CRBC}\|= \sqrt{\sum_{i=1}^{n} \mathbf{V}_{\mathrm{CRBC}_i}}$ and $\|\mathbf{V}_\mathrm{BC}\|= \sqrt{\sum_{i=1}^{n} \mathrm{V}_{\mathrm{BC}_i}}$.
The greater $\kappa(\mathbf{V}_\mathrm{CRBC}, \mathbf{V}_\mathrm{BC})$, the more relevant the code to be fixed and the bug-fix pair in the codebase is. At last, we design an RAG selection gate to ensure that only retrieved bug-fix pairs fulfilling the requirements are used as relevant bug-fix pairs. The details are shown below. 

\textbf{RAG Selection Gate.} 
As mentioned in \S~\ref{RAG}, we retrieve relevant bug-fix pairs based on semantics and AST similarity. The relevant information is used as a context for the model inputs to assist in repairing the target code. However, APR has a high requirement for efficiency and accuracy in real-world scenarios. If all retrieved bug-fixed pairs are added to the context, it may negatively affect the efficiency and accuracy of APR. On the one hand, it may cause the input sequence to be longer than the input length limit of the model, thereby incurring information loss due to truncation. On the other hand, if the extracted bug-fix pairs do not have a high enough degree of similarity with the target code, the added context will instead become a noisy input, degrading the accuracy. To address this challenge, we propose a selection gate mechanism. This process begins by using UniXcoder to encode and rank all retrieved information based on the similarity scores described previously. Given the token length limitations, we set a threshold for inclusion. Only bug-fix pairs with a similarity score exceeding this threshold are considered valid and added to the context. These selected pairs are then incorporated into the context in descending order of their similarity scores, until the token limit is reached. This approach ensures that the most relevant information is prioritized within the constrained context space. Therefore, the token length can be controlled while decreasing the inference time.

\subsection{APR Fine-tuning} \label{fine-tuning}
After selecting the relevant bug-fix pairs, we construct the input to fine-tune LLMs. The input consists of buggy codes from the training set and the retrieved valid bug-fix pairs. Inspired by~\cite{Wang2023}, we design a \emph{code-only} prompt supplementing a [BUG] token and a [FIX] token to concatenate buggy code and valid bug-fix pairs. The concatenated input  is shown as follows:
\begin{displaymath}
\small
\text{[BUG]} \ \text{RBC}_1\ \text{[FIX]}\ \text{RFC}_1\ \text{[BUG]} \ \text{RBC}_2\ \text{[FIX]}\ \textbackslash
\end{displaymath}
\begin{displaymath}
\small
\text{RFC}_2\ \ldots \ \text{RBC}_i\ \text{[FIX]}\ \text{RFC}_i\ \ldots \text{[BUG]}\ \text{BC}\ \text{[FIX]},
\end{displaymath}
where $\mathrm{RBC}_i$ represents the \underline{r}elevant \underline{b}uggy \underline{c}ode in the $i^\text{th}$ valid bug-fix pair, $\mathrm{RFC}_i$ represents the \underline{r}elevant \underline{f}ixed \underline{c}ode in the $i^\text{th}$ valid bug-fix pair, and $\mathrm{BC}$ represents the buggy code that needs to be repaired. The goal of the approach is to full-parameter fine-tune code LLMs to complete the fixed code in the above sequence. The objective function of fine-tuning is shown below:
\begin{equation}\small
   P_\theta(Y_{i}|X_{i}) = \prod \limits_{k=1}^n  P_\theta(y_{i,k}|X_{i},y_{i,1}, \ldots ,y_{i,k-1}),
\end{equation}
where $\theta$ is the parameter of the LLM, $X_{i}$ is $i^\text{th}$ the input sequence, $Y_{i}$ is the sequence with correct completed fixed code, and $y_{i,k}$ represents the $k^\text{th}$ token of the sequence with correct completed fixed code. The goal is to maximize the probability $P_\theta(Y_{i}|X_{i})$ by optimizing the parameter $\theta$. 

\begin{table*}
\centering
  \setlength{\abovecaptionskip}{-0mm}
  \caption{Evaluation Results for Compared Approaches}
  \label{comparison}
  \footnotesize
\renewcommand{\arraystretch}{0.8}
  \resizebox{0.8\linewidth}{!}{
  \begin{tabular}{lccccccccc}
    \toprule
    \multirow{2}{*}{\textbf{Approaches}} & \multicolumn{3}{c}{\textbf{Tufano Subset 1}} & \multicolumn{3}{c}{\textbf{Tufano Subset 2}} & \multicolumn{3}{c}{\textbf{VulRepair}}\\
    & \textbf{EM (\%)} & \textbf{BLEU-4} & \textbf{CodeBLEU} & \textbf{EM (\%)} & \textbf{BLEU-4} & \textbf{CodeBLEU} & \textbf{EM (\%)} & \textbf{BLEU-4} & \textbf{CodeBLEU}\\
    \midrule
    {\itshape \textbf {GPT-3.5}} & 2.58 & 11.67 & 56.78 & 1.72 & 12.38 & 63.64 &0.49&4.52&41.37\\
    {\itshape \textbf {GPT-4o}} & 0.17 & 6.24 & 57.46 & 0.00 & 7.19 & 59.68&0.00&2.14&30.95\\
    {\itshape \textbf {DeepSeek-R1-Distill}} & 0.09 & 3.24 & 37.52 & 0.00 & 2.47 & 34.68&0.00&0.83&18.84\\
    {\itshape \textbf {RAP-Gen}} & 24.80 & 69.77 & 76.33 & 15.84 & 85.27 & 85.92&23.02&48.20&51.67\\
    {\itshape \textbf {\system{}\textit{Llama}}} & 5.96 & 30.84 & 51.32 & 4.36 & 58.97 & 67.27 &6.82 & 28.42 & 39.37\\ 
    {\itshape \textbf {\system{}\textit{T5}}} & 25.27 & 65.98 & 76.57 & 16.36 & 80.23 & 84.81&24.36&43.83&58.85\\
    {\itshape \textbf {\system{}\textit{LoRA}}} & 22.62 & 57.46 & 72.99 & 13.05 & 74.06 & 82.19 & 0.73&34.98&49.94\\
    {\itshape \textbf {\system{}}} & \underline{\textbf{26.29}} & 61.61 & 74.35 &  \underline{\textbf{17.64}} & 73.88 & 82.24 &\underline{\textbf{25.46}} &38.39&50.84\\
    \bottomrule
  \end{tabular}
  }
\end{table*}

\subsection{Inference} \label{inference}
In the inference phase, we utilize test datasets to evaluate the performance of fine-tuned LLMs in generating patches. Specifically, we take each test sample and retrieve the relevant bug-fix pair via RAG as a context. We construct the same prompt as fine-tuning. In other words, we input a code-only prompt into fine-tuned LLMs to evaluate the generated patches using evaluation metrics. There are two kinds of test datasets: 1) the public code datasets and 2) another dataset coming from code fixes made by developers in a software-development enterprise during the development process. To simulate a real APR scenario, we use a search algorithm called beam search~\cite{Markus2017} to generate multiple patches for each test sample in the real-world test dataset to generate patches. With these datasets, we evaluate the performance of \system{} in both experimental and real scenarios.

\section{Experiments and Evaluation}

This section presents experiments to evaluate \system{}'s
performance in APR tasks and analyze the influencing factors of its performance. We aim to answer the following four research questions.
\begin{enumerate}[leftmargin=*,start=1,label={\bfseries RQ\arabic*:}]
\item What is the proposed \system{}'s performance compared with other state-of-the-art APR approaches?
\item What are the effects of different modules on \system's APR performance?
\item What are the effects of selection gate configuration on \system's APR performance?
\item What is \system{}'s performance in real-world scenarios?
\end{enumerate}

\subsection{Data Preparation \& Experiment Configurations}
\textbf{Datasets.}
In order to evaluate \system{}'s performance on program repair of different languages, we focus on Java and C/C++ program repair. Therefore, we conduct experiments on two Java datasets and a C/C++ dataset. We also introduce one additional dataset obtained from a software enterprise with an aim to evaluate the performance of \system{} in real-world scenarios. Appendix~\ref{sec:dataset} gives more details.

\textbf{Evaluation Metrics \& Experiments Configuration.} 
Three metrics are adopted to evaluate the APR performance: \underline{E}xact \underline{M}atch (\textbf{EM})~\cite{Armin2024}, 4-grams \underline{B}i\underline{l}ingual \underline{E}valuation \underline{U}nderstudy~\cite{Papineni2002} (\textbf{BLEU-4}) and \textbf{CodeBLEU}~\cite{ren2020} (refer to Appendix~\ref{metrics} for more details). The detailed hyperparameter settings are given in Appendix~\ref{configuration}.

\begin{table*}
\centering
  \setlength{\abovecaptionskip}{-0mm}
  \caption{Ablation Study}
  \label{ablation study}
  \renewcommand{\arraystretch}{0.8}
  \resizebox{0.8\linewidth}{!}{
  \begin{tabular}{lccccccccc}
    \toprule
    \multirow{2}{*}{\tabincell{c}{\textbf{Module}\\ \textbf{Construction}}} & \multicolumn{3}{c}{\textbf{Tufano Subset 1}} & \multicolumn{3}{c}{\textbf{Tufano Subset 2}} & \multicolumn{3}{c}{\textbf{VulRepair}}\\
    & \textbf{EM (\%)} & \textbf{BLEU-4} & \textbf{CodeBLEU} & \textbf{EM (\%)} & \textbf{BLEU-4} & \textbf{CodeBLEU} & \textbf{EM (\%)} & \textbf{BLEU-4} & \textbf{CodeBLEU}\\
    \midrule
    \textbf{w/o RAG \& Ft} & 0.00 & 7.96 & 36.96 & 0.00 & 3.32 & 31.71 & 0.00&8.05&28.26\\
    \textbf{w/o Ft} & 0.00 & 2.88 & 40.67 & 0.00 & 6.26 & 42.35 & 0.00 & 9.96 &31.17\\
    \textbf{w/o RAG} & 15.02 & 36.07 & 55.56 & 6.52 & 50.72 & 60.91 & 19.24 &37.58&50.46\\
    \textbf{w/o SR} & 25.57 & 58.16 & 73.94 & 10.83 & 60.40 & 71.18 & 21.92 & 36.79 & 50.06\\
    \textbf{w/o SSDR} & 22.28 & 56.81 & 72.98 & 17.41 & 73.73 & 82.12&22.68&38.28&50.64\\
    \itshape \textbf {\system{}} & \underline{\textbf{26.29}} & 61.61 & 74.35 &  \underline{\textbf{17.64}} & 73.88 & 82.24 &\underline{\textbf{25.46}} &38.39&50.84\\
    \bottomrule
  \end{tabular}
  }
\end{table*}

\subsection{RQ1: What is the proposed \system{}'s performance compared with other state-of-the-art
APR approaches?}
We compare \system{} with six state-of-art approaches, namely \textit{GPT-3.5}~\cite{Anis2023}, \textit{GPT-4o}~\cite{sun2024repofixeval}, \textit{DeepSeek-R1-Distill}~\cite{deepseekai2025}, \textit{RAP-Gen}~\cite{Wang2023}, \system{} with \textit{CodeLlama}-based LLM (\system{}\textit{Llama}), \system{} with \textit{CodeT5}-based LLM (\system{}\textit{T5}) and \system{} with LoRA fine-tuning (\system{}\textit{LoRA}). We describe the details of these approaches in Appendix~\ref{sec:baselines}. We choose these approaches because comparative approaches combine RAG with full-parameter fine-tuning and adopt code-only prompts similar to \system{}.
Moreover, we only consider an approach based on 
PEFT (i.e., LoRA). While we focus on approaches using RAG-based fine-tuning comparative to \system{}, we also include \textit{GPT-3.5}, GPT-4o, and \textit{DeepSeek-R1-Distill} in our comparison. These choices provide a baseline performance of a widely-used and general-purpose language model, thereby demonstrating the potential advantages of our approach in the APR context. The adoption of GPT-3.5 and GPT-4o also helps verify whether advanced generalized LLMs necessarily yield better APR performance. We leverage each approach to generate 1 repair candidate for each sample in the testing set.

The comparison results on Tufano's dataset are shown in Table~\ref{comparison}. The EM results of \textit{RAP-Gen} are obtained from the original paper while other metrics are experimentally evaluated by us. It can be found that \system{} achieves new SoTA performance of 26.29 EM and 17.64 EM in Tufano Subset 1 (< 50 tokens) and Tufano Subset 2 (50-100 tokens), respectively, outperforming other SoTA LLMs. Specifically, it outperforms \textit{GPT-3.5}, \textit{GPT-4o}, \textit{DeepSeek-R1-Distill}, \textit{RAP-Gen}, \system{}\textit{Llama}, \system{}\textit{T5}, and \system{}\textit{LoRA} by 918.99\%, 15364.71\%, 29111.11\%, 6.01\%, 341.11\%, 4.04\%, and 16.22\% respectively, in Tufano Subset 1. In Tufano Subset 2, \system{} performs 925.58\%, 11.36\%, 304.59\%, 7.82\% and 35.17\% better than \textit{GPT-3.5}, \textit{RAP-Gen}, \system{}\textit{Llama}, \system{}\textit{T5} and \system{}\textit{LoRA}, respectively. In summary, when inputting the code-only prompt, \system{} outperforms existing SoTA LLMs in terms of the percentage of correct repairs in both datasets with diverse code lengths. Moreover, we also observe that \system{}\textit{T5} outperforms 1.90\% than \textit{RAP-Gen} in Tufano Subset 1 and 5.28\% than \textit{RAP-Gen} in Tufano Subset 2. Since \system{}\textit{T5} and \textit{RAP-Gen} utilize the same base code LLM, the results indicate that \emph{the superiority of our design does not depend entirely on the scale of model parameters}. Other modules, including RAG and fine-tuning contribute to performance improvement. 
The exact contributions of RAG and fine-tuning are investigated in \S~\ref{RQ2}. Moreover, \system{}\textit{Llama} performs poorly (5.96 in Tufano Subset 1 and 4.36 in Tufano Subset 2), indicating that CodeLlama does not work well for code-only prompts. The mediocre performance in \system{}\textit{LoRA} indicates that full parameter fine-tuning contributes more than PEFT.

Besides Java datasets (Tufano), Table~\ref{comparison} also reports the comparison results on VulRepair (i.e., C/C++ dataset). Notably, we reproduce \textit{RAP-Gen} as well as other approaches on this dataset (\textit{RAP-Gen} did not adopt this dataset). It can be found that \system{} still achieves SOTA EM performance of 25.46, indicating its superior performance on different programming languages. 

We also observe that \textit{RAP-Gen} has a lower EM score than \system{} despite its slightly higher BLEU-4 and CodeBLEU scores. This implies that \textit{RAP-Gen} can generate results \textit{semantically more similar} to ground truth than our \system, but the correctness of the bug fixes generated by RAP-Gen may be less reliable than our \system, as indicated by its lower EM score. Moreover, we do not consider deep-learning-based approaches since \textit{RAP-Gen} is superior to most deep-learning-based approaches~\cite{Wang2023}. 
Further, \textit{GPT-based} models and \textit{DeepSeek-R1-Distill} have the worst performance, which may be attributed to the prompt design. As shown in Figure~\ref{GPT Prompt} in Appendix~\ref{sec:baselines}, we do not provide any bug type information or fine-grained bug location information (e.g., buggy line) for a fair comparison. Therefore, this kind of prompt cannot contribute to a good performance for general-purpose LLMs like \textit{GPT} and \textit{DeepSeek-R1-Distill}. We also find that \textit{GPT-4o} performs inferior to \textit{GPT-3.5}. This may be because \textit{GPT-4o} excels in general NLP tasks with complex data but performs poorly on some simple yet specific tasks.

\begin{table*}[t]
\centering
  \setlength{\abovecaptionskip}{-0mm}
  \caption{RAG Selection Gate Setting \& Efficiency Improvement}
  \label{threshold setting and improvement}
  \resizebox{\linewidth}{!}{
  \begin{tabular}{cccccccccccccccc}
    \toprule
    \multirow{3}{*}{\tabincell{c}{\textbf{Threshold}\\ \textbf{Setting}}}& \multicolumn{5}{c}{\textbf{Tufano Subset 1}} & \multicolumn{5}{c}{\textbf{Tufano Subset 2}} & \multicolumn{5}{c}{\textbf{VulRepair}}\\
    & \textbf{EM (\%)} & \textbf{BLEU-4} & \textbf{CodeBLEU} & {\tabincell{c}{\textbf{Avg. Input}\\ \textbf{Token Length}}} & {\tabincell{c}{\textbf{$\-$ Infer.}\\ \textbf{Time (\%)}}} & \textbf{EM (\%)} & \textbf{BLEU-4} & \textbf{CodeBLEU} &{\tabincell{c}{\textbf{Avg. Input}\\ \textbf{Token Length}}} & {\tabincell{c}{\textbf{$\-$ Infer.}\\ \textbf{Time (\%)}}} & \textbf{EM (\%)} & \textbf{BLEU-4} & \textbf{CodeBLEU} & {\tabincell{c}{\textbf{Avg. Input}\\ \textbf{Token Length}}} & {\tabincell{c}{\textbf{$\-$ Infer.}\\ \textbf{Time (\%)}}}\\
    \midrule
    \textbf{No Threshold} & 21.83 & 55.91 & 71.92 & 604.10 & - & 15.95 & 73.16 & 81.75 & 886.43 & - & 21.32 & 36.78 & 50.27 & 1175.09 & -\\
    \textbf{0.5} & 23.47 & 60.84 & 73.63 & 571.07 & 0.53 & 12.89&71.84&80.46 &867.42&7.79 &21.68&36.70&50.39&1162.36&4.49\\
    \textbf{0.7} & 23.73 & 57.67 & 73.67 & 68.24 & 2.27& 15.89 & 73.04 & 81.42& 169.22 & 8.94&23.02&38.00&50.13&1033.31&9.13\\
    \textbf{0.8} & 24.43 & 57.88 & 73.77 & 61.36 & 4.59 &\underline{\textbf{17.64}} & 73.88 & 82.24 & 133.16 & 13.77 &\underline{\textbf{25.46}} &38.39&50.84 &992.25&9.95\\
    \textbf{0.9} & \underline{\textbf{26.29}} & 61.61 & 74.35 & 60.53 & 6.42& 14.72 & 72.74 & 81.31& 131.05 & 29.68&23.75&38.90&51.93&986.70&15.96\\
    \bottomrule
  \end{tabular}
  }
  \vspace{-7.1mm}
\end{table*}

\subsection{RQ2: What are the effects of different modules on APR performance?} \label{RQ2}
As mentioned in \S~\ref{sec:method}, we adopt SR and SSDR as RAG and fine-tuning, respectively, to improve the APR performance. We design an ablation study to analyze how these modules contribute to the APR performance. We use ``without (w/o) RAG and Fine-tuning'', ``without (w/o) Fine-tuning'', ``without (w/o) RAG'', ``without (w/o) SR'', and ``without (w/o) SSDR'' as the module construction types and compare their performance with our baseline model. The results are shown in Table~\ref{ablation study}. It can be found that both fine-tuning and RAG (SR and SSDR) contribute to APR with different code lengths. \system{} outperforms ``w/o RAG'', ``w/o SR'', and ``w/o SSDR'' by 75.03\%, 2.82\%, and 18.00\%, respectively in Tufano Subset 1 (< 50 tokens). \system{} performs 170.55\%, 62.88\%, and 1.32\% better than ``w/o RAG'', ``w/o SR'', and ``w/o SSDR'' in terms of EM in Tufano Subset 2 (50-100 tokens). In Tufano Subset 1, SSDR has a more significant contribution because the static structure and dependency information of AST is more useful in short code to complement the semantic and syntax information, thereby helping the model to better understand the syntax and semantics of the code. Further, \system{} outperforms ``w/o RAG'', ``w/o SR'', and ``w/o SSDR'' by 32.33\%, 16.15\%, and 12.26\%, respectively in VulRepair. The results indicate that SR, SSDR and RAG all contribute to APR performance in different programming languages. \system{} also has the best BLEU-4 and CodeBLEU scores among all the methods. Notably, ``w/o RAG and Fine-tuning'' and ``w/o Fine-tuning'' achieve an EM score of 0.00 in different datasets, suggesting that Code LLMs struggle to interpret code-only prompts without fine-tuning.


\subsection{RQ3: What are the effects of selection gate configuration on APR performance?} \label{RQ3}
To find the optimal setting for the RAG selection gate, we design an experiment to analyze the effect of different selection gate threshold settings (0.9, 0.8, 0.7, 0.5, and No Threshold). Table~\ref{threshold setting and improvement} reports the results, showing that \system{} has the best performance in Tufano Subset 1 (< 50 tokens) when the threshold is 0.9. In Tufano Subset 2 (50-100 tokens) and VulRepair, \system{} has the best performance when the threshold setting is 0.8. The results indicate that too much RAG information may not enhance \system{}'s performance.

We also analyze the efficiency of different threshold settings. We observe from Table~\ref{threshold setting and improvement} that both the input token length and the inference time decrease with the increased threshold. When the threshold value is 0.9, the inference time is 6.42\%, 29.68\%, and 15.96\% less than no threshold setting in Tufano Subset 1 (< 50 tokens), Tufano Subset 2 (50-100 tokens) and VulRepair, respectively. Therefore, a suitable threshold setting not only improves the performance but also keeps the inference time within an acceptable range. This finding also has implications for the design of other RAG-based tasks. In other words, the RAG selection gate can also be adapted to other RAG-based LLM tasks. The acceleration of inference enhances its reliability in industrial practice.

Considering the trade-off between inference time and APR performance, we set 0.9 as the default threshold for Tufano Subset 1 and 0.8 for Tufano Subset 2 and VulRepair.

\subsection{RQ4: What is \system{}'s performance in real-world scenarios?}
\begin{figure}
  \centering
  \setlength{\abovecaptionskip}{-0.5mm}
  \includegraphics[width=0.9\linewidth]{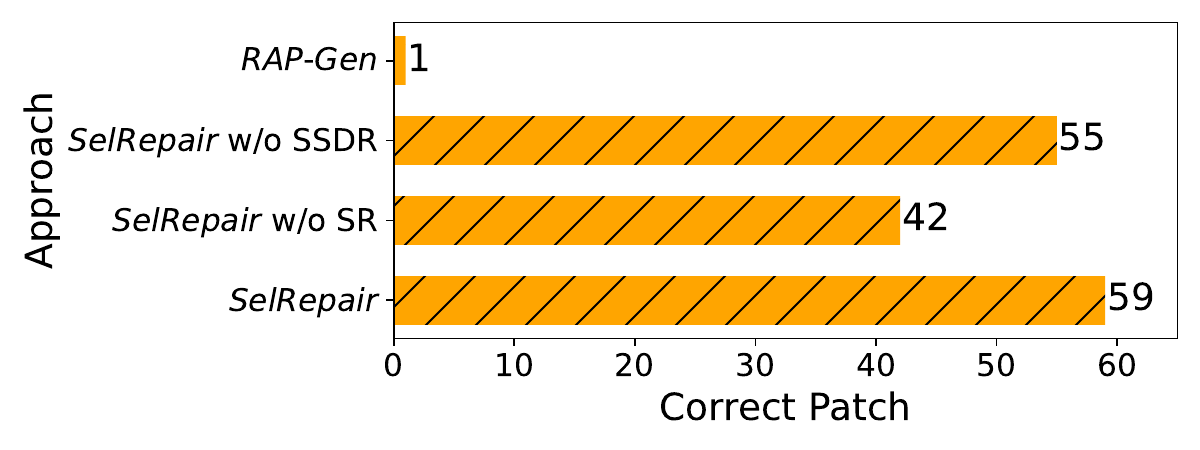}
  \caption{Performance on Real-world Enterprise Data}
  \label{enterprise}
  \vspace{-5.8mm}
\end{figure}

We also adopt a benchmark of 200 bug-fix pairs from an enterprise to verify the performance of \system{} in real-world scenarios. This enterprise benchmark differs from open-source benchmarks like the Tufano dataset. While the Tufano dataset primarily addresses functional defects, the enterprise benchmark emphasizes coding bad practices and style issues, such as improper logging, unnecessary checks, and unused variables. The enterprise benchmark is designed to align with organizational coding standards. We intend to open-source this benchmark in the future, potentially providing a new resource for APR research. We count the number of correct patches for \system{}, \system{} w/o SR, \system{} w/o SSDR, and \textit{RAP-Gen} (fine-tuned with Tufano). The results are shown in Figure~\ref{enterprise}. As for beam search, we set the beam size to 10, indicating that each sample can have 10 generated patches. It can be found that \system{} also achieves the best performance among all the approaches by generating 59 correct patches. When we adopt either SSDR or SR, \system{} can still generate 42 and 55 correct patches. In contrast, \textit{RAP-Gen} fine-tuned with the Tufano dataset can only generate 1 correct patch for our benchmark. The results demonstrate the excellent \textit{generalizability} of \system{}.

\textbf{Other discussions.} In Appendix~\ref{sec:discussions}, we present a discussion of how \system{} works, give a generated case study, and analyze \system{}'s performance on \textit{Defects4J}. In Appendix~\ref{sec:threats}, we present the threats to validity of \system{}.

\vspace{-2.5mm}
\section{Conclusion}
\vspace{-2.5mm}

In this paper, we present \system{}, an innovative APR approach leveraging fine-tuned LLMs with a dual RAG strategy. By fully fine-tuning LLMs using bug-fix pair datasets, we tailor the model to effectively address APR challenges. Our dual RAG module incorporates semantic, syntactic, and structural information, overcoming the limitations of current RAG mechanisms that focus solely on semantics. Additionally, we design an RAG selection gate to verify its role in the repair process. Our evaluation on three open datasets and one enterprise dataset shows \system{} outperforms state-of-the-art methods, enhancing APR effectiveness and efficiency. Future work includes extending to other programming languages and integrating real-time feedback mechanisms to improve accuracy across diverse environments.

\section*{Limitations}
Our current method is limited by datasets focused on individual methods, which simplifies research but misses real-world complexity. Many bugs result from interactions across methods or modules, requiring comprehensive analysis. Future work should expand techniques to handle larger code spans while preserving semantic integrity and control flows for holistic debugging.

The large parameter size of current state-of-the-art LLMs poses additional challenges by requiring significant resources for fine-tuning and limiting accessibility. Integrating these models into DevOps environments with quick response needs is difficult. While compression methods like distillation and quantization offer solutions, they can affect performance. Balancing expressiveness with deployment requirements remains challenging.

To overcome these limitations, we are exploring several research directions. We are developing techniques for cross-method and cross-component bug handling by integrating structural information. We're exploring efficient fine-tuning methods like meta-learning to optimize resource use. For deployment, we're modularizing models and enhancing caching and indexing to reduce latency, aiming to improve LLM-based debugging in real-world software development.


\bibliography{reference}

\appendix

\section{Taxonomy of Adopting LLMs} \label{sec:taxonomy}

\begin{figure*}[t]
  \centering
  \includegraphics[width=\linewidth]{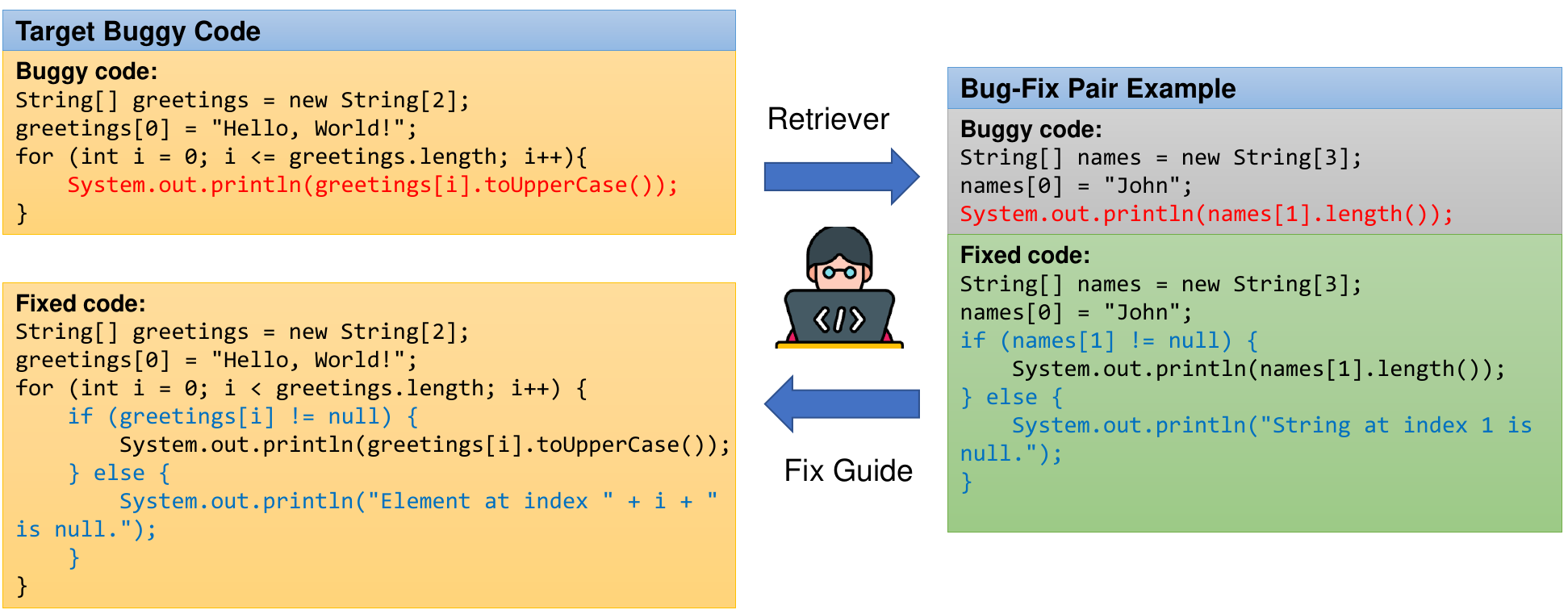}
  \caption{An Example of RAG in APR}
  \label{RAG example}
\end{figure*}

\subsection{Prompt engineering}
\textbf{Prompt engineering} refers to the design and optimization of input prompts to obtain the best output from an LLM~\cite{white2023}. It does not require extra training but its performance depends on pre-training tasks. Since most popular generalized LLMs (e.g., GPT-3.5, GPT-4 etc.~\cite{ye2023}) and code LLMs (e.g., CodeT5~\cite{wang2021}, CodeBERT~\cite{feng2020}, StarCoder~\cite{li2023}, StarCoder 2~\cite{lozhkov2024} etc.) do not include APR-related pre-training tasks, it is difficult to design an ideal set of prompts to target generic APR tasks. 

\subsection{Fine-tuning}
\textbf{Fine-tuning} is the use of task-specific data (e.g., APR data) to further train a model based on an LLM~\cite{lin2024}. Despite many fine-tuned code pre-trained models for APR-related tasks, most of them are adopted on LLMs with less than 1B parameters~\cite{Ehsan2021, Huang2023}. Although other approaches fine-tune LLMs with more than 1B parameters~\cite{yang2024}, they adopt Parameter-Efficient Fine-Tuning (PEFT) techniques~\cite{melnyk23a} (e.g., LoRA~\cite{hu2022lora}, adaptor tuning~\cite{houlsby19a,silva2024repairllama}) rather than full-parameter fine-tuning~\cite{lv2023}. As a result, they cannot fully unleash the potential of LLMs in APR.

\section{A Motivation Example of RAG} \label{sec:RAG example}
Figure~\ref{RAG example} shows an example to describe how RAG contributes to APR. In the target buggy code, it contains a null pointer exception (NPE) bug since an attempt is made in a \texttt{\small for} loop to access an element in an array of strings that may not have been initialized (i.e, \texttt{\small greetings[1]}). By using the RAG, the APR module can retrieve an NPE-related bug-fix pair example. In this bug-fix pair, the key to the fix is to check if an array element is \texttt{\small null} before attempting to access it. The LLM then uses this bug-fix pair as a context to generate fixed code for the original bug, i.e., adding a \texttt{\small null} check.

\section{Hybrid Retriever Algorithm}\label{app:hybrid}
Algorithm~\ref{algHR} depicts how the hybrid reviewer algorithm works. 
\begin{algorithm}[h]
\SetKwInput{KwInput}{Input}
	\SetKwInput{KwOutput}{Output}
 	\SetKwComment{Comment}{//}{}
    \SetKwFunction{Function}{\emph{hybrid\_retriever($C$,$T$,$t$)}}
	\caption{Hybrid Retriever}
	\label{algHR}
    \footnotesize
	\KwInput{\footnotesize ${C}$: Bug-fix pairs; \footnotesize ${T}$: Target buggy code; \newline
	    \footnotesize ${t}$: Similarity threshold.}
 \KwOutput{\footnotesize $\mathrm{BF}$: Retrieved bug-fix pair set}
 
        \textbf{function} \Function
               
        $\mathrm{BF}$ $\gets$ [\emph{$\emptyset$}]

        $\mathrm{AST}_T$ = \texttt{AST\_Parse(${T}$)}\label{line:ASTParse}

        $\mathrm{ASTSeq}_T$ = \texttt{AST\_traversal($\mathrm{AST}_T$)}\label{line:ASTtraversal}

        $\mathbf{V}_\mathrm{BCs}$ = \texttt{UnixCoder($T$)}\label{line:BCc}

        $\mathbf{V}_\mathrm{BCa}$ = \texttt{UnixCoder($\mathrm{ASTSeq}_{T}$)}\label{line:BCa}

        $\mathbf{V}_\mathrm{BC}$ = ($\mathbf{V}_\mathrm{BCs}$ + $\mathbf{V}_\mathrm{BCa}$)/2
        
        \For{\emph{$\mathrm{CRBC}$, $\mathrm{CRFC}$} in ${C}$} 
        {       \label{line:iterate}
        $\mathrm{AST}_\mathrm{BF}$ = \texttt{AST\_Parse($\mathrm{CRBC}$)} \label{line:ASTParseCR}
        
        $\mathrm{ASTSeq}_\mathrm{BF}$ = \texttt{AST\_traversal($\mathrm{AST}_\mathrm{BF}$)} \label{line:ASTtraversalCR}

        $\mathbf{V}_\mathrm{CRBCs}$ = \texttt{UnixCoder($\mathrm{CRBC}$)} \label{line:CRBCc}

        $\mathbf{V}_\mathrm{CRBCa}$ = \texttt{UnixCoder($\mathrm{ASTSeq}_\mathrm{BF}$)} \label{line:CRBCa}

        $\mathbf{V}_\mathrm{CRBC}$ = ($\mathbf{V}_\mathrm{CRBCs}$ + $\mathbf{V}_\mathrm{CRBCa}$)/2 \label{line:CRBC}

        \If {$\kappa(\mathbf{V}_\mathrm{CRBC}, \mathbf{V}_\mathrm{BC})$ > $t$} 
        {   \label{line:CosSimilarity}
        $\mathrm{BF}$.\textit{append}((\emph{$\mathrm{CRBC}$, $\mathrm{CRFC}$}))}  
        }

        \KwRet {${BF}$}
\end{algorithm}

\section{Details of AST traversal} \label{sec:AST traversal}
\begin{figure}[!htbp]
  \centering
  \includegraphics[width=\linewidth]{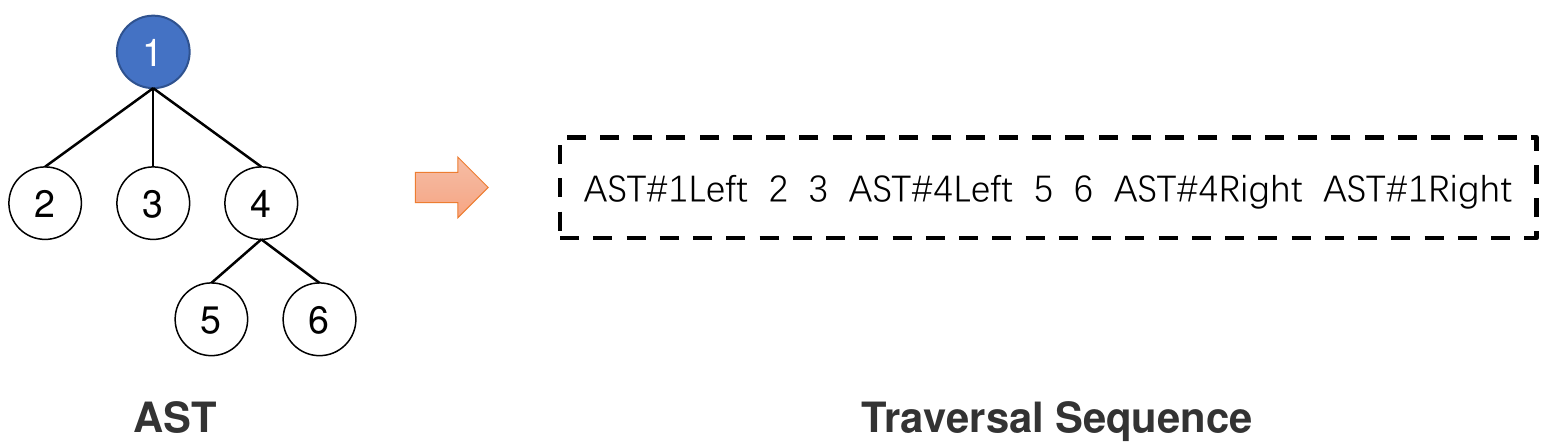}
  \caption{A Toy Example of AST Traversal}
  \label{AST traversal}
\end{figure}

Algorithm~\ref{traversal} depicts the procedure of AST traversal, in which we add nodes to the sequence in the order of pre-order traversal. If the node is a leaf node, the node \textit{value} information is added to the sequence directly (lines~\ref{line:ifleaf}-\ref{line:append}). If the node is a non-leaf node, it will transform to \texttt{AST\#node\_type\#left} and \texttt{AST\#node\_type\#right} tokens, while the information of its corresponding child nodes is added between these two tokens (lines~\ref{line:else}-\ref{line:elseappend}). Figure~\ref{AST traversal} shows a toy example of AST and the corresponding AST sequence.
\begin{algorithm}[t]
\SetKwInput{KwInput}{Input}
	\SetKwInput{KwOutput}{Output}
 	\SetKwComment{Comment}{//}{}
    \SetKwFunction{Function}{\emph{AST\_traversal($R$)}}
	\caption{AST Traversal}
	\label{traversal}
    \footnotesize
	\KwInput{\footnotesize ${R}$: The root node of the AST;} 
	
 \KwOutput{\footnotesize ${S}$: The traversed AST sequence;}
 
        \textbf{function} \Function
        
        ${S}$ $\gets$ [\emph{$\emptyset$}]

        \eIf {${R}$ is \emph{leaf\_node}}
        { \label{line:ifleaf}
        ${S}$.\textit{append}(${R}$.\textit{value})\label{line:append}
        } 
        {\label{line:else}
        ${S}$.\textit{append}(`AST\#'+${R}$.\textit{type}+`\#Left')
        
        \For{\emph{node} in ${R}$.\textit{children}}
        {${S}$.\textit{extend}(\texttt{AST\_traversal}(node))}
        ${S}$.\textit{append}(`AST\#'+${R}$.\textit{type}+`\#Right') \label{line:elseappend}
        }

        \KwRet {${S}$}
\end{algorithm}

\section{Details of Experiment Setup}
\subsection{Details of Dataset Construction} \label{sec:dataset}

We consider two Java datasets, a C/C++ dataset and a software enterprise's Java dataset to evaluate the performance of \system{}.

We firstly evaluate \system{} on a public dataset proposed by Tufano et al.~\cite{Tufano2019}. It consists of bug-fix pairs at the method level and it is collected from fix commit records from GitHub. Specifically, it contains two data subsets of split according to the length of the code token. One is the subset with code lengths of less than 50 tokens (i.e., < 50 tokens dataset), and the other is the subset with code lengths of 50-100 tokens (i.e., 50-100 tokens dataset). These two subsets are named Tufano Subset 1 and Tufano Subset 2. The distribution of these two subsets is shown in Table~\ref{data distribution}. As for each subset, we random sample 1,000 samples as an RAG codebase. For the remaining samples, we split 80\% of the dataset as a training set, 10\% as a validation set, and 10\% as a test set.

Another dataset is a C/C++ dataset proposed by Fu et al.~\cite{Fu2022}, which is called VulRepair. It consists of bug-fix pairs combined by CVE-Fixes~\cite{Guru2021} and Big-Vul~\cite{Fan2020}. We filtered out invalid samples, such as samples that were null. The distribution of this datsaet is also shown in Table~\ref{data distribution}. Similarly, we randomly sample 2000 samples as an RAG codebase. For the remaining samples, we split the dataset into training set, validation set, and a test set in a ratio of 8:1:1.
\begin{table*}
\centering
  \caption{Distribution of Dataset}
  \label{data distribution}
  \begin{tabular}{lcccccc}
    \toprule
    \textbf{Datasets} & \textbf{Language} & \textbf{Code Length}&\textbf{RAG Codebase}&\textbf{Train}&\textbf{Valid}&\textbf{Test}\\
    \midrule
    Tufano Subset 1& Java & < 50 tokens & 1,000&45,880&5,735&5,735\\
    Tufano Subset 2& Java & 50-100 tokens& 1,000& 51,565&6,447&6,447\\
    VulRepair & C/C++ & - & 200 & 6,574 & 822 & 821\\
  \bottomrule
\end{tabular}
\end{table*}

In order to evaluate the performance of \system{} in real scenarios, we also introduce one additional dataset, which comes from a software enterprise. This dataset consists of 200 semantic bug-fix pairs caused by enterprise developers in real development scenarios. We verify the effectiveness of \system{} to fix errors in realistic scenarios by using this dataset.

\begin{figure*}[h]
  \centering
  \includegraphics[width=\linewidth]{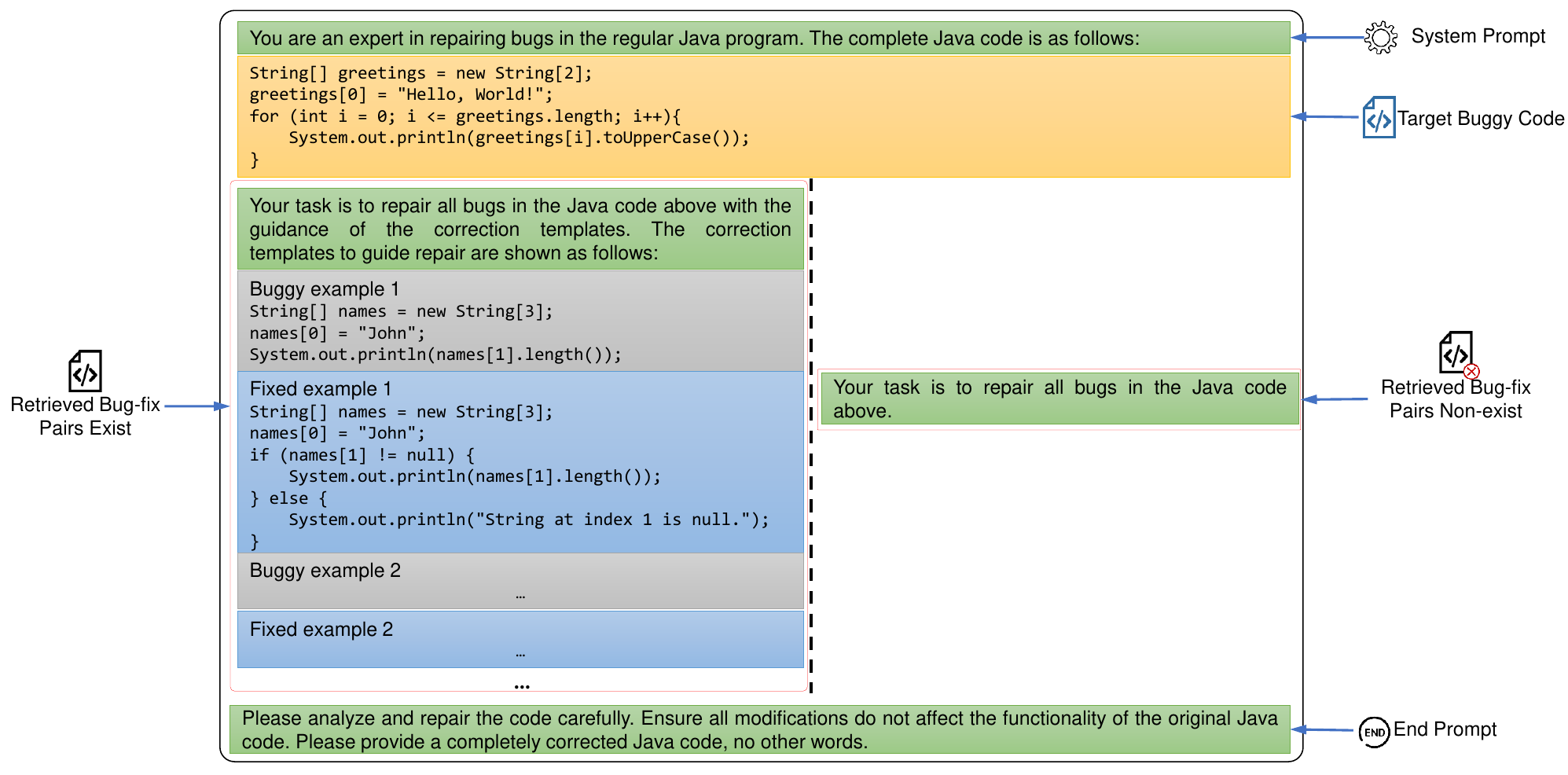}
  \caption{GPT-3.5 \& GPT-4o Prompt Template}
  \label{GPT Prompt}
\end{figure*}

\subsection{Evaluation Metrics}\label{metrics}
We adopt EM, BLEU-4, and CodeBLEU to evaluate the APR performance.
\begin{itemize}[leftmargin=*]
\item \textbf{EM} refers to the ratio of generated fixes identical to the ground truth made by developers (i.e., reference fixes). Although there may be multiple fixes for the same bug, it can be used as an indicator of the performance of fixing logic bugs. 

\item \textbf{BLEU-4} is a commonly used machine translation evaluation metric that measures the similarity between the predicted text and the reference text. We utilize BLEU-4 as a looser metric to evaluate the similarity between generated fixes and reference fixes. It first splits the generated fix and the reference fix into 1-gram to 4-grams. Then, for each $n$-gram (1 to 4), BLEU-4 calculates the number of overlaps between the n-gram in the generated fix and the $n$-gram in the reference fix, as well as a weighted geometric mean of the 1-gram to 4-grams precision. The specific calculation process of BLEU-4 is given as follows:
\begin{equation}\label{BLEU-4}
  \operatorname{BLEU4} = \mathrm{BP} \cdot \exp(\sum_{n=1}^{4}\omega_{n} \log p_{n}),
\end{equation}
where $\omega_{n}$ is the weight of $n$-grams, $p_{n}$ is the precision of $n$-gram, and $\mathrm{BP}$ refers to the brevity penalty factor for the generated fix length. $\mathrm{BP}$ is given as follows: 
\begin{equation}
\mathrm{BP}=\left\{
\begin{aligned}
1 \quad\quad & , & f_{g}> f_{r}, \\
\exp\bigg(1-\frac{f_{r}}{f_{g}}\bigg) & , & f_{g}\leq f_{r},
\end{aligned}
\right.
\end{equation}
where $f_{g}$ is the length of generated fix and $f_{r}$ represents the reference fix.

\item \textbf{CodeBLEU} is a code-specific evaluation metric derived from BLEU. It enables the quality assessment of APR tasks while preserving BLEU's benefits through $n$-gram matching and injecting code syntax and semantics through ASTs and data flows. CodeBLEU is calculated as follows:
\begin{equation} 
\begin{aligned}
\operatorname{CodeBLEU} = \alpha \cdot \operatorname{BLEU} + \beta \cdot \operatorname{BLEU}_\text{weight} + \\
\gamma \cdot \operatorname{Match}_\text{ast} + \epsilon \cdot \operatorname{Match}_\text{df},
\end{aligned}
\end{equation}
where $\operatorname{BLEU}$ is the standard BLEU calculated by Eq.~\eqref{BLEU-4} ($\omega_{1}$ to $\omega_{4}$ are all equivalent). $\operatorname{BLEU}_\text{weight}$ refers to the weighted $n$-gram match calculated by Eq.~\eqref{BLEU-4} ($\omega_{1}$ to $\omega_{4}$ can be different). $\operatorname{Match}_\text{ast}$ refers to syntactic AST match, addressing the syntactic information of code. $\operatorname{Match}_\text{ast}$ is calculated as follows:
\begin{equation}
  \operatorname{Match}_\text{ast} = \frac{\mathrm{Count_{clip}}(\text{ST}_\text{gen})}{\mathrm{Count}(\text{ST}_\text{ref})},
\end{equation}
where $\operatorname{Count}(\text{ST}_\text{ref})$ refers to the total number of the subtrees of ASTs parsed from reference fixes, and $\operatorname{Count}_\text{clip}(ST_\text{gen})$ is the number of the subtrees of ASTs parsed from generated fixes that are matched the reference. $\operatorname{Match}_\text{df}$ refers to the semantic data-flow match score, which is calculated as follows:
\begin{equation}
  \operatorname{Match}_\text{df} = \frac{\mathrm{Count_{clip}}(\text{DF}_\text{gen})}{\mathrm{Count}(\text{DF}_\text{ref})},
\end{equation}
where $\operatorname{Count}(\text{DF}_\text{ref})$ is the total number of the reference fixes' data flows, and $\operatorname{Count}_\text{clip}(\text{DF}_\text{gen})$ is the number of matched data-flows from generated fixes. $\alpha$, $\beta$, $\gamma$ and $\epsilon$ are weight coefficients designed by the user.
\end{itemize}

\subsection{Experiment Configuration} \label{configuration}
The hyperparameter setting is shown as follows. Referring to~\cite{wang2024-1}, we set the fine-tuning epochs as 3 for the large parameter (> 1B) LLM. We set the context window as 512 tokens for the Tufano Subset 1 (< 50 tokens), 1,024 tokens for the Tufano Subset 2 (50-100 tokens) and 1,500 tokens for VulRepair dataset. We adopt \textit{StarCoder2-7B} as the foundation code LLM for fine-tuning. As for the optimizer, we utilize Adam~\cite{Kingma2015} with the learning rate $5 \times 10^{-5}$ for supervised fine-tuning (SFT). The threshold of RAG selection gate is set as 0.9 for Tufano Subset 1 and 0.8 for Tufano Subset 2 and VulRepair dataset. More details are shown in \S~\ref{RQ3}. All experiments are conducted on a server configured with 4 GPUs of NVIDIA GeForce RTX 3090.

\begin{figure*}[!htbp]
  \centering
  \includegraphics[width=\linewidth]{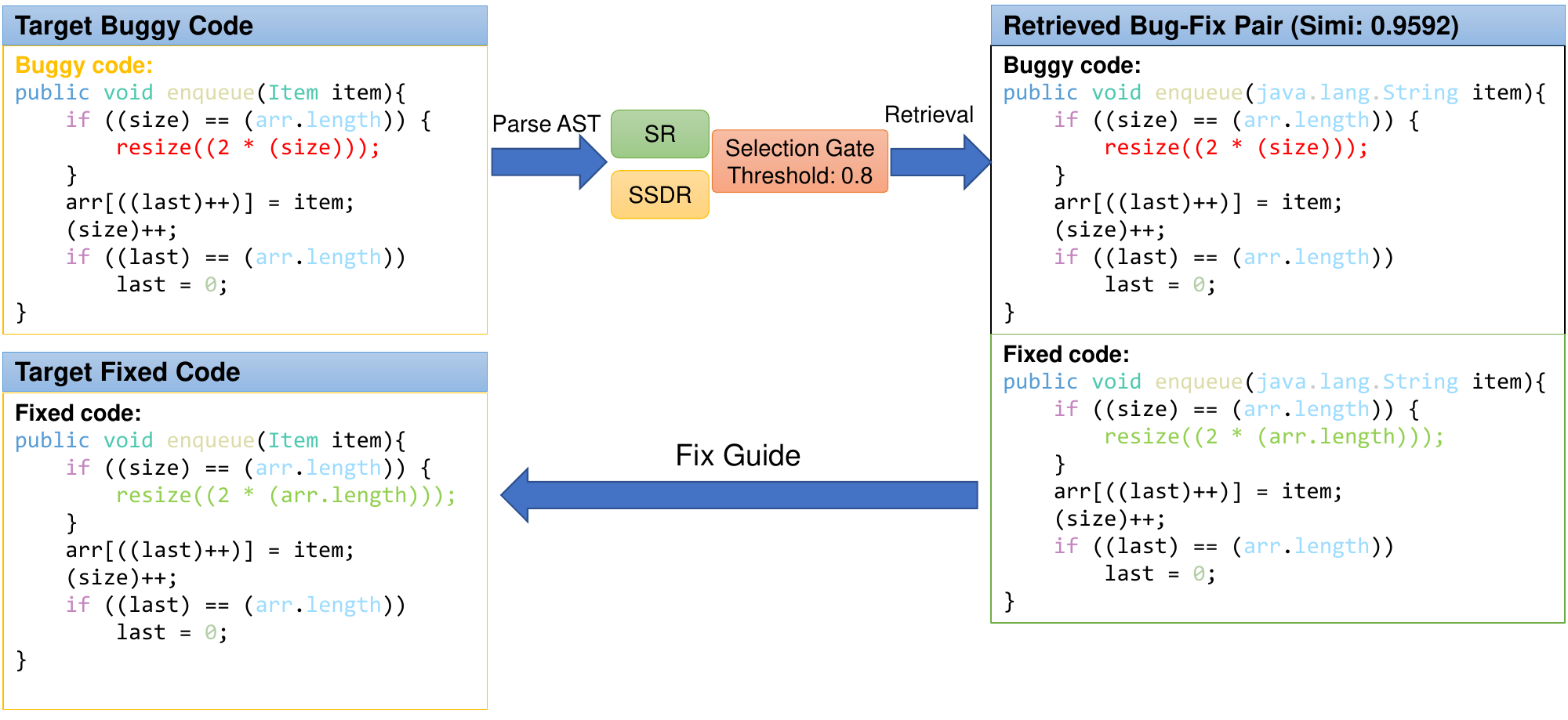}
  \caption{Detailed Process of \system{}}
  \label{DetailedProcess}
\end{figure*}

\subsection{Baselines} \label{sec:baselines}
We adopt four state-of-the-art approaches as the baselines to compare with \system{}, which are shown as follows: 
\begin{itemize}[leftmargin=*]
    \item \textit{\textbf{GPT-3.5}}: \textit{GPT-3.5} is a General-purpose Large Language Model developed by OpenAI that offer significant architectural and performance improvements compared with previous LLMs. It is based on the Transformer architecture. Since \textit{GPT-3.5} is a General-purpose Large Model, referring to~\cite{xu2024}, we design an instruction-based prompt to implement the APR task, as shown in Figure~\ref{GPT Prompt}. It includes system prompt and target buggy code. If retrieved bug-fix pairs exist, we add them to the prompt. Otherwise, we directly tell the model to perform the APR task via instructions. Finally, we add an end prompt to ask the model to generate the fixed code. For a fair comparison, the prompt does not contain a description of the bug type and bug location information. The utilization of \textit{GPT-3.5} aims to measure whether \system{} outperforms APR methods by adopting instruction-based prompt engineering.
     \item \textbf{\textit{GPT-4o}}: GPT-4o is one of the latest LLMs developed by OpenAI, and it has been significantly improved and enhanced in several aspects compared to GPT-3.5, including larger parameter sizes, and more training data, as well as support for multimodal inputs and outputs. We design the same instruction-based prompt as GPT-3.5 to implement the APR task.

     \item \textbf{\textit{DeepSeek-R1-Distill}}: \textit{DeepSeek-R1} is a general-purpose inference model developed by DeepSeek AI company. \textit{DeepSeek-R1} uses reinforcement learning for post-training and is designed to improve inference, and is particularly adept at complex tasks such as mathematical, coding, and natural language reasoning. \textit{DeepSeek-R1-Distill} models are fine-tuned based on open-source models, using samples generated by \textit{DeepSeek-R1}. For a fair comparision, we adopt a 7B-parameter \textit{DeepSeek-R1-Distill} model, that is, \textit{DeepSeek-R1-Distill-Qwen-7B}. It is fine-tuned based on \textit{Qwen2.5}~\cite{qwen2.5-VL} LLM.
    \item \textbf{\textit{RAP-Gen}}: This approach adopt fine-tuning on \textit{CodeT5} and semantics similarity as RAG. As mentioned in~\cite{Wang2023}, it outperforms most popular deep-learning-based APR approaches and code-LLM-based approaches. So, we adopt \textit{RAP-Gen} as one of SoTA LLMs.
    \item \textbf{\system{}\textit{Llama}}: In Appendix~\ref{configuration}, we adopt \textit{StarCoder2-7B} as the foundation code LLM. We also consider another earlier code LLM called \textit{CodeLlama} as the foundation code LLM. We aim to compare the performance of code LLMs released at different times on the target task. We also adopt three fine-tuning epochs for this approach.
    \item \textbf{\system{}\textit{T5}}: To verify that our approach also improves in code LLM with small-scale parameters, we replace the foundation code LLM with \textit{CodeT5}. Referring to the design of \textit{RAP-Gen}, we adopt 50 fine-tuning epochs for this approach.
    \item \textbf{\system{}\textit{LoRA}}: Considering PEFT-based methods, we try to adopt LoRA fine-tuning strategy for \system{}.  LoRA (Low-Rank Adaptation)~\cite{hu2021loralowrankadaptationlarge} is an approach for fine-tuning LLMs. It enables efficient fine-tuning by adjusting some of the weights of the model without significantly increasing the number of parameters. We also adopt three fine-tuning epochs for this approach.
\end{itemize}

\section{Discussions} \label{sec:discussions}

\subsection{Detailed Process of How \system{} Works}
We present a real buggy code snippet in the test set and show how \system{} fixes this buggy code. The detailed fixing process is shown in Figure~\ref{DetailedProcess}. The size variable in line 3 needs to be replaced with variable \textcolor{blue}{\texttt{arr.length}} to ensure the rationalization of array expansion. When \system{} receives the buggy code, the code will be parsed into AST. The code and AST will be input to the SR and SSDR module to get the feature vector and calculate the similarity with each sample in the codebase. Then we adopt the selection gate and set the similarity threshold to 0.8. A bug-fix pair in the codebase is retrieved and the similarity is 0.9592. The bug-fix pair provides a similar fix pattern so that \system{} can fix the bug with the use of this RAG information.

\begin{figure*}[!ht]
  \centering
  \includegraphics[width=0.55\linewidth]{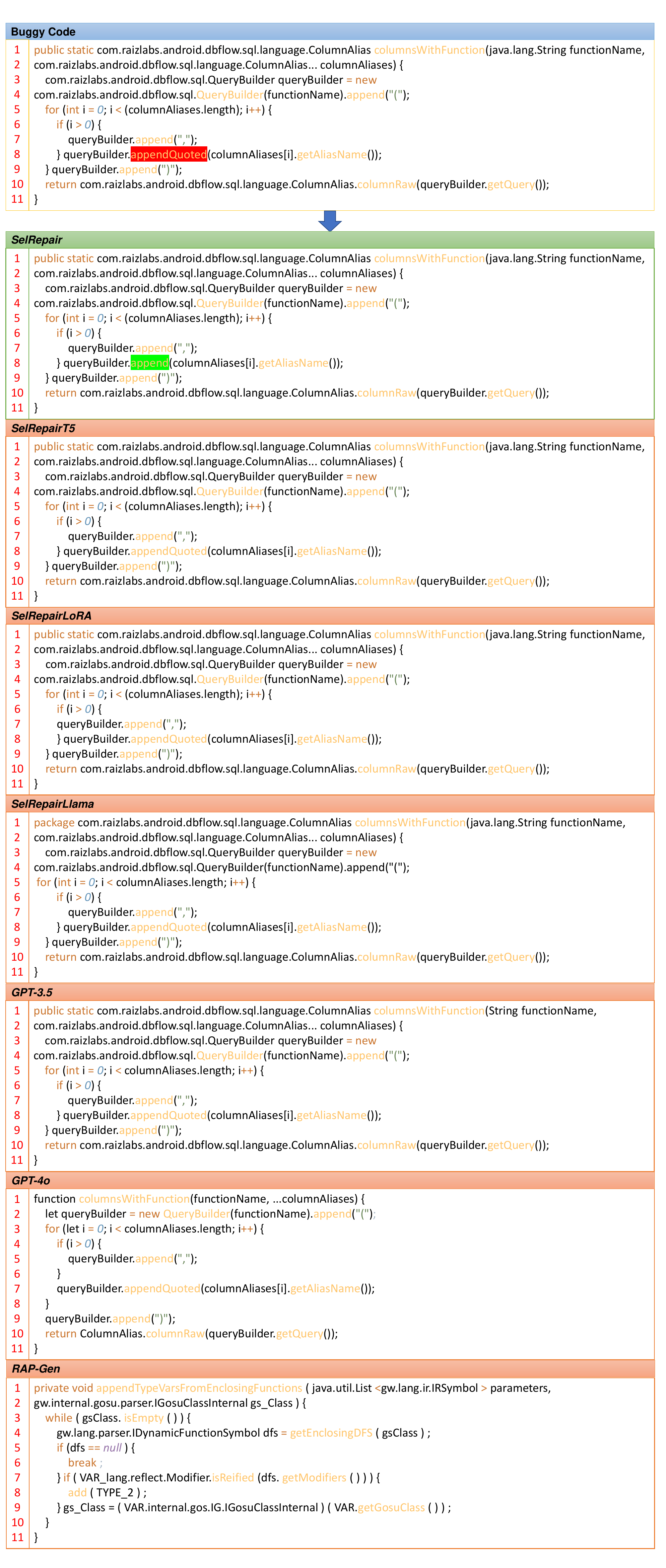}
  \caption{Case Study}
  \label{case2}
\end{figure*}

\subsection{Case Study}
In this section, we propose a patch case generated by the \system{} and other SOTAs. Figure~\ref{case2} presents an example from Tufano Subset 2 (50-100 tokens). The buggy code at line 8 incorrectly uses the \textcolor{blue}{\texttt{appendQuoted}} method instead of \textcolor{blue}{\texttt{append}}. The key difference between them is as follows:

\begin{itemize}[leftmargin=*]
    \item \textcolor{blue}{\texttt{append}}: This method is used to append a string value to the existing content of a StringBuilder object without adding any quotation marks.
    \item \textcolor{blue}{\texttt{appendQuoted}}: This method is specifically designed to append a string value to the StringBuilder object while enclosing the value in quotation marks.
\end{itemize}

In the given context, using \textcolor{blue}{\texttt{append}} is the appropriate choice since the \textcolor{blue}{\texttt{getAliasName()}} method already returns the column name without quotation marks. Using \textcolor{blue}{\texttt{appendQuoted}} may result in extra quotation marks being added around the column name, leading to incorrect syntax. Among the compared approaches, only \system{} successfully generates the correct patch for this bug. In contrast, \system{}\textit{T5} and \system{}\textit{LoRA} generate the same code as buggy code. \system{}\textit{Llama} changes
\textcolor{blue}{\texttt{public static}} to \textcolor{blue}{\texttt{package}}, which indicates that this approach misunderstands method \textcolor{blue}{\texttt{fcolumnsWithFunction}}.
\textit{GPT-3.5} makes an invalid modification that change the type of \textcolor{blue}{\texttt{functionName}} from \textcolor{blue}{\texttt{java.lang.String}} to \textcolor{blue}{\texttt{String}}. \textit{GPT-4o} also makes the invalid modification (change \textcolor{blue}{\texttt{public static}} to \textcolor{blue}{\texttt{function}}. As for \textit{DeepSeek-R1-Distill}, we find that it outputs excessively long reasoning content, suffers from over-reasoning, and does not generate the repaired code at the end, which may indicate that the model is impaired in comprehending this code. Therefore, we do not present the content generated by \textit{DeepSeek-R1-Distill}. Considering \textit{RAP-Gen}, it cannot comprehend the semantics of the code and generate the wrong patch.
 
The case highlights \system{}'s ability to generate accurate patches for code implementation errors that cannot be detected by static analysis tools. Such errors often require a deeper understanding of code semantics and the intended functionality. By accurately fixing these implementation errors, \system{} demonstrates its robustness and effectiveness in program repair tasks. It showcases the model's ability to comprehend the nuances of code semantics and generate patches that align with the intended functionality, even when errors are not detectable by traditional static analysis tools.

\begin{table*}
\centering
  \caption{Performance on \textit{Defects4J}}
  \label{Defects4J}
  \resizebox{0.9\linewidth}{!}{
  \begin{tabular}{ccc}
    \toprule
    \textbf{Approaches} & \textbf{Defects4J V1.2} & \textbf{Defects4J V2.0} \\
    \midrule
    \textbf{\system{} (Beam Size = 10)} & 35 & 11 \\
    \textbf{\textit{RAP-Gen} (Beam Size = 10)} & 32 & 12 \\
    \textbf{\textit{GPT-3.5}(10 Generated Patches)} & 11 & 3 \\
  \bottomrule
\end{tabular}
}
\end{table*}

\subsection{Performance on \textit{Defects4J}}
\textit{Defects4J}~\cite{Just2014} is one of the most widely adopted APR datasets. Based on our collation of its two versions (v1.2 and v2.0), \textit{Defects4J} contains 1,273 bug-fix pairs at the method level from 17 open-source Java projects on GitHub. As mention in~\cite{Wang2023}, \textit{RAP-Gen} adopts a project-specific training data curated by SelfAPR~\cite{He2023} and evaluate \textit{Defects4J}. Specifically, \textit{RAP-Gen} is trained with a dataset constructed by the same projects as \textit{Defects4J}. This configuration may cause data leaks and weaken the generalizability of the approach. Therefore, referring to~\cite{Huang2023}, we utilize a dataset proposed by Jiang et al.~\cite{Jiang2023} to fine-tune the \system{} and test the performance on \textit{Defects4J}. Besides, we adopt beam search and set the beam size as 10. For a fair comparison, we set \textit{GPT-3.5} to generate 10 patches for each bug. We count the number of patches that can pass the test cases. The results are shown in Table~\ref{Defects4J}. It can be found that \system{} can generate 35 patches and 11 patches for v1.2 and v2.0, respectively. As for \textit{RAP-Gen} fine-tuned with the data from the same projects, it can generate 32 patches for v1.2 and 12 patches for v2.0. \system{} outperforms RAP-Gen at a beam size of 10 in Defects4J V1.2 and achieves a close performance to RAP-Gen in Defects4J V2.0. In general, \system{} has better performance on cross-project APR data. In particular, we do not consider some other prompt-engineering-based approaches such as~\cite{li2024}. Although this kind of approach is orthogonal to \system{} and may have better results for simple APR tasks, for complex tasks, it is not possible to extend beyond the original capabilities of the LLMs due to the dependence on the capabilities of the pre-trained models, so our approach extends the complex APR capabilities through fine-tuning, which is based on the addition of prompt engineering.

\section{Threats to Validity} \label{sec:threats}
The threats to validity include internal validity, external validity and construct validity.

\textbf{Internal validity} addresses the correctness and reliability of our experiments and data processing. Issues can arise from errors in the bug-fix dataset and biases during language model fine-tuning, such as overfitting. To mitigate these, we implemented rigorous data preprocessing and validation steps. Another concern is the threshold settings in the RAG selection gate, where coarse-grained thresholds were used for different code lengths. Future work will focus on automatically setting customized thresholds for diverse code lengths.

\textbf{External validity} concerns whether our findings extend beyond the Java and C/C++ datasets used. While \system{} showed promise in repairing Java programs, its effectiveness with other languages like Python or JavaScript is untested. Language syntax, semantics, and bug patterns may impact performance. Future work will involve evaluating diverse datasets from multiple languages to assess and refine \system{}'s adaptability, ensuring broader applicability and robustness across different software development environments.

\textbf{Construct validity} ensures our metrics and benchmarks accurately reflect program repair effectiveness. We plan to evaluate our approach on diverse datasets from different lengths to ensure generalizability and compare results with established benchmarks and other methods. Testing in real-world environments will assess practical applicability. Developer feedback will provide insights into perceived utility and accuracy. So far, we have used open-source data for training and testing. We also use an internal enterprise dataset to ensure broader applicability. These steps will strengthen construct validity by ensuring accurate and applicable performance across contexts.

\section{Data Availability}
We make our approach available at \url{https://anonymous.4open.science/r/SelRepair-5F1D/}.

\end{document}